\documentclass[aps,prb,amsfonts,amssymb,twocolumn,amsmath,preprintnumbers,nofootinbib,floatfix,
showpacs]{revtex4-1}
\usepackage[dvips]{graphics}
\usepackage{graphicx}
\usepackage{bm}
\begin{document}

\title{Recovering the superconducting state of superconductor-ferromagnet
    multilayer structures by spin accumulation above the Pauli pair-breaking
    magnetic field}
\author{I. V. Bobkova}
\affiliation{Institute of Solid State Physics, Chernogolovka,
Moscow reg., 142432 Russia}
\author{A. M. Bobkov}
\affiliation{Institute of Solid State Physics, Chernogolovka,
Moscow reg., 142432 Russia}

\date{\today}

\begin{abstract}
We study theoretically the simultaneous influence of spin accumulation potential $eV_\uparrow-eV_\downarrow$
and the Zeeman exchange field on singlet superconductivity. It is shown that the pair-breaking effect
of the Zeeman field can be fully compensated by creation of the appropriate spin accumulation potential
in the superconductor. Moreover, superconductivity can be recovered for exchange fields well exceeding the
Pauli limiting field. It is proposed that the effect can be experimentally realized on the basis of voltage
biased junction consisting of a thin superconducting film sandwiched between two half metals. 
\end{abstract}
\pacs{74.78.Fk, 74.40.Gh}

\maketitle

One of the mechanisms destroying singlet superconductivity 
is the Zeeman interaction of electron spins with magnetic field. The behavior of a magnetic superconductor
with an exchange field $h$ 
was studied long ago \cite{larkin64,fulde64,sarma63,maki68}. It was found that 
homogeneous superconducting state becomes energetically unfavorable above the paramagnetic (Pauli) limit
$h=\Delta/\sqrt 2$. As it was predicted \cite{larkin64,fulde64}, in a narrow region of exchange 
fields exceeding this value superconductivity can appear as an inhomogeneous state with a spatially 
modulated Cooper pair wave function (LOFF-state). Thus, the exchange field is destructive to singlet superconductivity. 
An exception, proposed in the literature, is a special type of clean superconducting 
multilayered system, where the paramagnetic limit can be enhanced and a nonuniform superconducting 
state can be induced under in-plane magnetic field \cite{buzdin05}. 

Superconductor/ferromagnet (S/F) hybrid structures also can behave analogous to magnetic superconductors. In particular,
it was shown \cite{bergeret01} that a thin S/F bilayer with thicknesses obeying the conditions $d_F \ll \xi_F$ 
and $d_S \ll \xi_S$ is equivalent to a magnetic superconductor with an effective exchange field 
$h_{eff}=h N_F d_F/(N_F d_F + N_S d_S)$ and an effective superconducting order parameter $\Delta_{eff}=
\Delta N_S d_S/(N_F d_F + N_S d_S)$. Here $\xi_F=\sqrt{D_F/h}$ and 
$\xi_S=\sqrt{D_S/|\Delta|}$ are the magnetic and superconducting coherence lengths, 
$|\Delta|$ is the superconducting order parameter in the bulk material, $N_{F,S}$ denote the densities
of states at the Fermi level for the ferromagnet and superconductor and $D_{F,S}$ are
the corresponding diffusion constants. Another way to "apply" an exchange field on a thin superconducting
film is to contact it to a ferromagnetic insulator \cite{tedrow86,meservey94,moodera88,hao91,cottet09}. 
It was observed experimentally \cite{hao91} and justified theoretically \cite{cottet09} that 
the effective exchange field induced in the film scales with $d_S^{-1}$. 

The simultaneous applying of the exchange field and creation of spin-dependent quasiparticle 
distribution in S/F heterostructures can lead to qualitatively new phenomena. In particular, it was shown recently 
\cite{bobkova10,bobkov11} that creation of spin-dependent quasiparticle distribution in the interlayer 
of S/F/S junction (or S/N/S junction with magnetic S/N interfaces) leads to appearence of an additional 
contribution to the Josephson current through the junction, which under certain conditions can enhance it
considerably. In the present paper we demonstrate that for a thin superconducting film the destructive effect
of the exchange field can be fully compensated by the creation of spin-dependent quasiparticle 
distribution in it. This effect takes place even if the exchange field exceeds the paramagnetic limit considerably,
that is under the condition that superconductivity of the
equilibrium film is fully suppressed.   

As an example of a system, where the uniform exchange field and the spin-dependent quasiparticle
distribution can be realized simultaneously, we propose here a voltage-biased 
half metal/superconductor/half metal (HM/S/HM) heterostructure. 
A thin film ($d_S \ll \xi_S$) made of a dirty s-wave superconducting 
material is sandwiched between two half-metallic layers with opposite directions of magnetization. 
Half-metallic behavior has been reported in ${\rm CrO}_2$ \cite{soulen98,ji01} 
and in certain manganites \cite{park98}. In-plane effective uniform exchange field $h_{eff}$ 
in the film is supposed to be created by spin-active interfaces with half metals. The spin-dependent 
quasiparticle distribution in the film can be generated by applying a voltage bias between the two half metals.
In this case for spin-up subband the main voltage drop occurs at one of the HM/S interfaces, while for spin-down subband - 
at the other. As a result, the distribution functions for spin-up and spin-down electrons in the superconducting
film are to be close to the equilibrium form with different electrochemical potentials. 

As we consider a non-equilibrium system, we make use of Keldysh framework of the quasiclassical theory, where
the fundamental quantity is the momentum average of the quasiclassical Green's function 
$\check g(x,\varepsilon) = \langle \check g(\bm p_f, x ,\varepsilon) \rangle_{\bm p_f}$. Here $x$
is the coordinate normal to S/HM interfaces and $\varepsilon$ is a quasiparticle energy. The Green's function
is a $8\times8$ matrix form in the product space of Keldysh,
particle-hole and spin variables. In the superconducting film it obeys the Usadel equation
\begin{equation}
\frac{D}{\pi} \partial_x (\check g \partial_x \check g)+\left[ \varepsilon \tau_3 \sigma_0 \rho_0
- \Delta \tau_1 i \sigma_2 \rho_0 , \check g \right]=0
\label{usadel}
\enspace ,
\end{equation}
where $\tau_i$, $\sigma_i$ and $\rho_i$ are Pauli matrices in particle-hole, spin and 
Keldysh spaces, respectively. $\tau_0$, $\sigma_0$ and $\rho_0$ stand for the corresponding 
identity matrices. Eq.~(\ref{usadel}) should be supplied with the normalization
condition $\check g^2 =-\pi^2\tau_0\sigma_0\rho_0$. It is convenient to express Keldysh part of the full Green's function via 
the retarded and advanced components and the distribution function: 
$\check g^K=\check g^R \check \varphi-\check \varphi \check g^A$. The distribution function is diagonal in particle-hole
space: $\check \varphi=\hat \varphi (\tau_0+\tau_3)/2+ \sigma_2 \hat {\tilde \varphi} \sigma_2
(\tau_0-\tau_3)/2$. The hole component $\hat {\tilde \varphi}$
of the distribution function is connected to $\hat \varphi$ by general symmetry relation \cite{serene83}
$\hat {\tilde \varphi}(\varepsilon)=-\sigma_2 \hat \varphi (-\varepsilon) \sigma_2$. 

We consider the case when there is the only magnetization axis in the system (the magnetization
directions of the PM's are antiparallel). Then there are no equal-spin
triplet superconducting correlations in the system and all the matrices
in spin space can be represented as sums of two spin subband contributions ($\sigma=\uparrow,\downarrow$). 
For later use we define here the anomalous Green's function 
$f^{R,A}=(f^{R,A}_\uparrow (\sigma_0+\sigma_3)/2+f^{R,A}_\downarrow (\sigma_0-\sigma_3)/2)i\sigma_2$ 
and the distribution function $\hat \varphi = \varphi_\uparrow 
(\sigma_0+\sigma_3)/2+\varphi_\downarrow (\sigma_0-\sigma_3)/2$.

The Usadel equation is subject to appropriate boundary conditions at S/HM interfaces, which for the diffusive limit
can be written in the form \cite{cottet09}
\begin{eqnarray}
\check g \partial_x \check g = -\alpha \frac{G_T^{l,r}}{2\sigma_S}\left[ \check g, \check g_{HM}^{l,r} \right]-
~~~~~~~~~~~~~~~\nonumber \\
-\alpha \frac{G_{MR}^{l,r}}{2\sigma_S}\left[ \check g, \left\{ \check m^{l,r}, \check g_{HM}^{l,r} \right\}\right]+
\alpha \frac{G_\phi^{l,r} \pi}{2\sigma_S}\left[ \check m^{l,r}, \check g \right]
\label{boundary_magnetic}
\enspace ,
\end{eqnarray}
where $\check g$ is the Green's function value at the superconducting side of the appropriate S/HM interface
(at $x=\mp d_S/2$), $\alpha=+1(-1)$ at the left (right) S/HM interface and $\sigma_S$ stands for the conductivity
of the film. $\check m^{l,r}=\bm m^{l,r}
\bm \sigma \rho_0 (1+\tau_3)/2+\bm m^{l,r} \bm \sigma^* \rho_0 (1-\tau_3)/2$, where $\bm m^{l,r}$
is the unit vector aligned with the magnetization direction of the left or right half metal. 
We assume that the half metals have opposite magnetization directions, that is $\bm m^r=-\bm m^l$. 
The second term accounts for the different conductances of different spin directions and 
$G_{MR} \sim G_{T,\uparrow}-G_{T,\downarrow}$.
The third term $\sim G_{\phi}$ gives rise to spin-dependent phase shifts of quasiparticles being reflected at
the interface. Microscopically \cite{cottet09} $G_\phi^{l,r} = 2 (G_q /S) \sum_n 
(T_n^{l,r}-1) d\phi_n^{l,r} $, where $S$ is the junction area and $G_q=e^2/h$ is the quantum conductance. 
Summation over $n$ means summation over transmission channels. $T_n^{l,r}$ is the transmission probability 
for the $n^{{\rm th}}$ channel and  $d \phi_n^{l(r)}$  is the phase difference betweeen wave functions of
spin-up and spin-down electrons, acquired upon reflection from S/HM interface (spin-dependent phase shift). Boundary 
conditions (\ref{boundary_magnetic}) are only valid for small (with respect to unity) values of transparency 
$T_n$ and spin-dependent phase shift $d \phi_n$ in one transmission
channel.  The value of $d \phi_n^{l(r)}$  can be roughly estimated, for example, 
by modelling the barrier at S/HM interface by
$U_\sigma(x)=U_\sigma \delta(x)$. Then in the tunnel limit $T_n \ll 1$ one obtains 
$d \phi_n \approx v_F(U_\downarrow-U_\uparrow)/U_\uparrow U_\downarrow$. In general, the boundary conditions can contain another term proportional to $G_\chi$ accounting
for spin-dependent phase shifts of quasiparticles upon transmission \cite{cottet09}. However, we are mostly
interested in the tunnel limit, where this term can be disregarded with respect to $G_\phi$. 

$\check g_{HM}^{l,r}$ stands for the Green's functions at the half metallic side of the interface.
Since in half metals a Fermi surface only exists for one of the
spin orientations, the standard quasiclassical description is inapplicable. However, half metals
still allow for a straightforward quasiclassical treatment
in the separate-band picture: quasiparticle trajectories
simply exist only for one of the spin orientations \cite{grein09}. 
If one chooses the quantization axis along the left HM magnetization, 
$\check g_{HM}^{l,r}$ take the form 
\begin{equation}
\check g_{HM}^{R,A}=-i\pi \kappa \tau_3 (\sigma_0+\alpha \sigma_3)/2
\label{green_bulk}
\enspace .
\end{equation}
Here labels $(l,r)$ are omitted for brevity and $\kappa = +1 (-1)$ for the retarded (advanced) Green's functions.
The distribution functions in the half metals are assumed to have the 
equilibrium form shifted by the applied voltages $V_{l,r}$. We suppose that $V_r=-V_l=V$. In this case
\begin{eqnarray}
\check g_{HM}^K=-2i \pi \left[\tanh \frac{\varepsilon+\alpha eV}{2T}\frac{(\tau_0+\tau_3)}{2}-
\right. 
\nonumber \\
\left. 
\tanh \frac{\varepsilon-\alpha eV}{2T}\frac{(\tau_0-\tau_3)}{2}\right]\frac{\sigma_0+\alpha\sigma_3}{2} 
\label{keldysh_bulk}
\enspace .
\end{eqnarray}
  
The self-consistent order parameter in the film is expressed via the Keldysh part of the anomalous 
Green's function. We assume that the paring constant is non-zero only for the singlet pairing channel.
Then the corresponding self-consistency equation takes the form 
\begin{eqnarray}
\Delta = \frac{\lambda}{2} \int \limits_{-\omega_D}^{\omega_D} \frac{d \varepsilon}{4 \pi i} \sum \limits_\sigma
\left[ f^R_\sigma(\varepsilon) \tilde \varphi_\sigma(\varepsilon)+ f^A_\sigma(\varepsilon) \varphi_\sigma
(\varepsilon)\right]
\label{self_con_distrib}
\enspace ,
\end{eqnarray}
where $\lambda$ is the dimensionless coupling constant. 

Eqs.~(\ref{usadel})-(\ref{self_con_distrib}) constitute a full system of equations for solving the problem. For the case
of a thin superconducting film $d_S \ll \xi_S$ they can be solved analytically just analogously to the case
of thin S/F bilayer \cite{bergeret01}, mentioned in the introduction. Averaging retarded and advanced parts
of Eq.~(\ref{usadel}) over the thickness of the film and taking into account the boundary conditions, one can reduce 
the Usadel equation to an equation describing an uniform magnetic superconductor with
an effective exchange energy $h_{eff}$ and a decoherence factor $\Gamma$.
This equation can be easily solved. The corresponding anomalous components of retarded and advanced Green's
function, entering Eq.~(\ref{self_con_distrib}) take the form
\begin{equation}
f_\sigma^{R,A}=\frac{\pi \Delta}{\sqrt{\Delta^2-\left[ \varepsilon+\sigma h_{eff}+i\kappa \Gamma \right]^2}}
\label{anomalous}
\enspace ,
\end{equation}
where the decoherence factor $\Gamma=(G_T^r+2G_{MR}^r+G_T^l+2G_{MR}^l)D/4\sigma_S d_S$ physically describes
the leakage of superconducting correlations from the film into the HM regions. This term is quite standard 
(including S/N systems) and is responsible for the destroying of superconductivity in thin films by the proximity
effect.  

The effective exchange energy $h_{eff}=(G_\phi^r-G_\phi^l)D/2\sigma_S d_S$ is generated by S/HM interfaces.
It is inversely proportional to the film width $d_S$. As it was mentioned above,
the boundary conditions (\ref{boundary_magnetic}) and, correspondingly, Eq.~(\ref{anomalous}) are
valid for $ d \phi_n \ll 1$. Beyond this limit the effect of magnetic boundaries cannot be reduced to
the effective exchange in the film \cite{cottet09}, but there appear additional terms to some extent analogous
to the magnetic impurities. We assume that the condition $ d \phi_n \ll 1$ is fulfilled. However, this does not
mean that the resulting exchange fields are small. In order to observe recovering of superconductivity,
suppressed by the exchange field, one needs $h_{eff} \gtrsim \Delta$. For the film with $d_S \lesssim \xi_S$
this condition is accomplished if (i) the left and right interfaces are not identical, that is $G_\phi^r \neq G_\phi^l$
and (ii) the dimensionless parameter $h_{eff}/\Delta \sim G_\phi \xi_S^2 / \sigma_S d_S \gtrsim 1$.
This parameter can be estimated as $-(2 N \xi_S^2 G_q/S \sigma_S d_S) d \phi 
\sim -(\xi_S^2/l d_S) d \phi$, where $N$ is the number of transmission channels and $l$ is the mean free path. 
From this estimate it is seen that
for dirty superconductors $G_\phi$ can generate large $h_{eff}$ even for $ d \phi \ll 1$.

\begin{figure}[!tbh]
  \begin{minipage}[b]{0.5\linewidth}
     \centerline{\includegraphics[clip=true,width=1.5in]{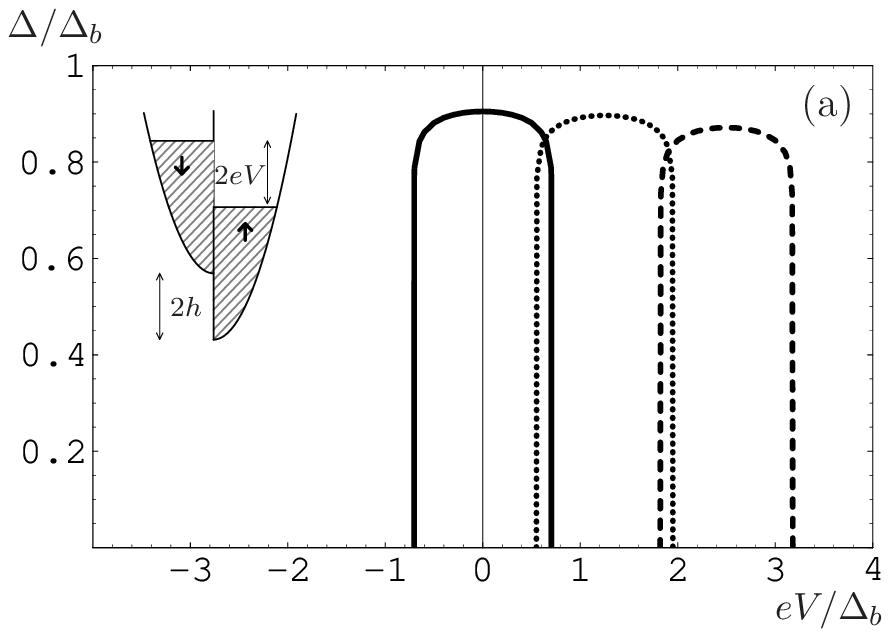}}
     \end{minipage}\hfill
    \begin{minipage}[b]{0.5\linewidth}
   \centerline{\includegraphics[clip=true,width=1.5in]{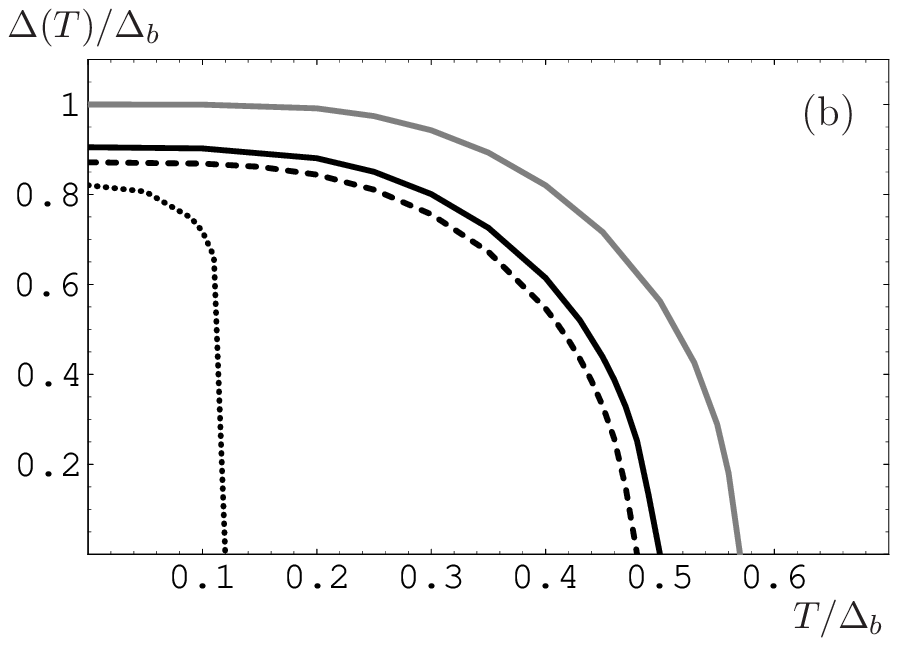}}
  \end{minipage}
\begin{minipage}[b]{0.5\linewidth}
     \centerline{\includegraphics[clip=true,width=1.5in]{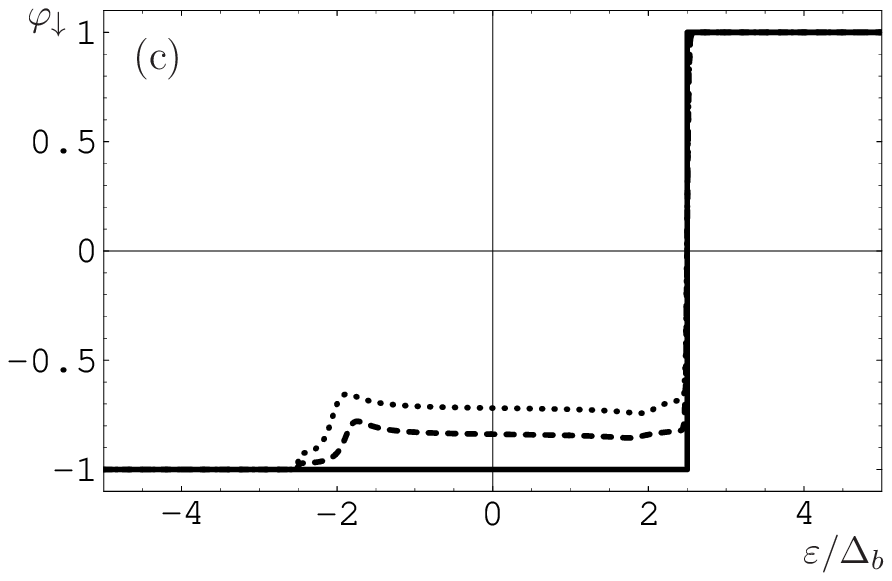}}
     \end{minipage}\hfill
    \begin{minipage}[b]{0.5\linewidth}
   \centerline{\includegraphics[clip=true,width=1.5in]{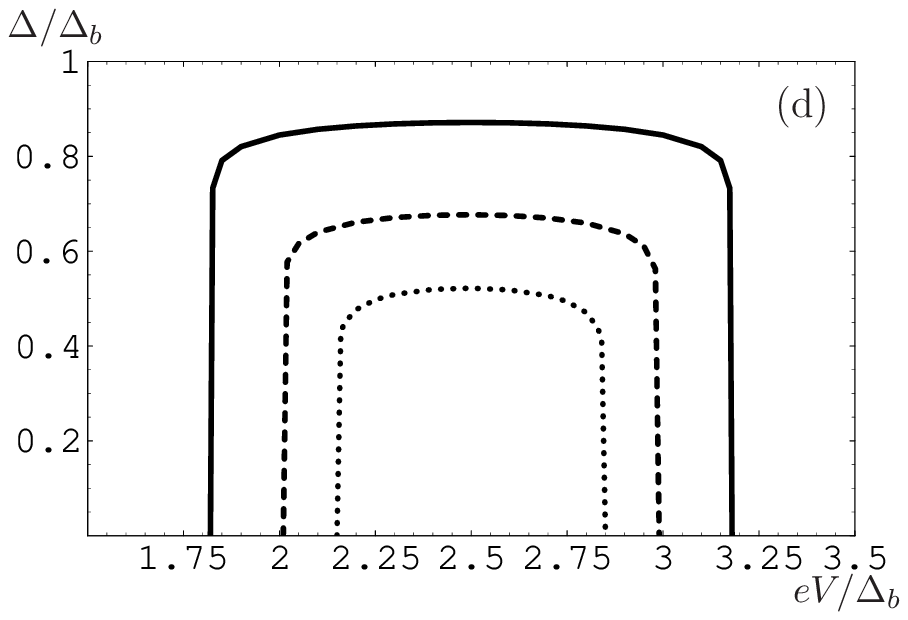}}
  \end{minipage}
   \caption{(a) The dependence of $\Delta(T=0)$ on $eV$. Solid line: $h_{eff}=0$, dotted: $h_{eff}=1.25$, dashed: 
$h_{eff}=2.5$. All the quantities are measured in units of the bulk order parameter $\Delta_b$ 
taken at $1/\tau_{sf}=0$ and $T=0$. Inset:
population of majority and minority subbands for the given quasiparticle distribution.
(b) The dependence of $\Delta$ on temperature. Gray solid line: $\Delta_b(T)$, black solid: $\Delta_0(T)
\equiv \Delta(T, h_{eff}=eV=0)$, dashed: $\Delta(T,h_{eff}=eV=2.5)$, dotted: $\Delta(T, h_{eff}=2.5, eV=3.1)$. 
For panels (a) and (b) $1/\tau_{sf}=0$. (c) The distribution function $\varphi_\downarrow(\varepsilon)$ 
for different spin relaxation rates: $1/\tau_{sf}=0$ (solid line), $0.01$ (dashed)
and $0.02$ (dotted). $h_{eff}=eV=2.5$. $\varphi_\uparrow(\varepsilon)=-\varphi_\downarrow(-\varepsilon)$. 
(d) The dependence of $\Delta$ on $eV$ for $h_{eff}=2.5$ and different spin relaxation rates (the same 
as in panel (c)). For all the panels $\Gamma=0.1$.}   
\label{fig}
\end{figure}

Now let us turn to the discussion of the distribution function. We neglect energy relaxation in the film,
that is assume that the time $\tau_{esc}=\Gamma^{-1}$, which an electron spends in the film is much less than the energy
relaxation time $\tau_\varepsilon$. Spin relaxation processes are also not taken into account. We discuss their
influence below. Then it can be obtained from Eqs.~(\ref{usadel})-(\ref{keldysh_bulk}) that
the distribution function in the film takes the form
\begin{equation}
\varphi_\sigma = \tanh \frac{\varepsilon+\sigma eV}{2T}
\label{distrib}
\enspace .
\end{equation}
It is worth noting here that the distribution function has such a one-step shape
(in each of the spin susbbands) due to the fact that the leads are HM: the electrons from spin-up (spin-down)
subband can flow only to/from the left (right) lead. This one-step form is very essential 
for the existence of the effect. In principle, the superconductivity recovering can be also observed
if one takes strong ferromagnets instead of half metals, but in this case the nonequilibrium distribution 
function inside the film is represented by a sum of the distribution functions coming from 
the left and right leads, weighted by factors depending on the interface transparencies (this is a double-step
structure). This would lead to only partial recovering of superconductivity, or even to the absence of the effect.
So, in order to provide the appropriate distribution function in the film the resistances of the ferromagnets
and S/F interfaces should obey quite strict conditions.

Substituting Eqs.~(\ref{anomalous})-(\ref{distrib}) [$\tilde \varphi$ is obtained 
making use of the symmetry relation $\tilde \varphi_{\uparrow,\downarrow}(\varepsilon)=
-\varphi_{\downarrow,\uparrow}(-\varepsilon)$] into Eq.~(\ref{self_con_distrib}) we come to the 
following self-consistency equation
\begin{eqnarray}
\frac{1}{\lambda} = \int \limits_{-\omega_D}^{\omega_D} \frac{d \varepsilon}{4 } \left\{{\rm Re}
\left[ \frac{{\rm sgn}(\varepsilon + h_{eff})}{\sqrt{(\varepsilon+i\Gamma+h_{eff})^2-\Delta^2}}\right]\tanh 
\frac{\varepsilon+eV}{2T} 
\right. 
\nonumber \\
\left. + {\rm Re}
\left[ \frac{{\rm sgn}(\varepsilon - h_{eff})}{\sqrt{(\varepsilon+i\Gamma-h_{eff})^2-\Delta^2}}\right]\tanh 
\frac{\varepsilon-eV}{2T} \right\}. ~~~~~~~~~~ 
\label{self_con_BCS}
\end{eqnarray}

From Eq.~(\ref{self_con_BCS}) it is obvious how superconductivity in the 
film is recovered under the simultaneous influence of the exchange field and the spin-dependent
quasiparticle distribution. At $h_{eff}=eV$ for each of the subbands we have practically the same 
situation as for the equilibrium non-magnetic film corresponding to $eV=h_{eff}=0$. It is worth noting here
that in the framework of the simplified weak-coupling model with constant pairing potential 
the maximal value of the Zeeman field, which does not destroy the superconductivity, equals to the 
cutoff energy $\omega_D$. However, for high enough exchange fields this 
simplified model is inapplicable. For concreteness let us discuss the phonon-mediated superconductivity. 
In this case our consideration fails for exchange fields of the order of the Debye energy 
and the correct calculation should be carried out in the framework of the particular phonon model. 
 
The resulting $\Delta$ as a function of applied voltage $eV$ is plotted in panel (a) of Fig.~\ref{fig}
for different values of $h_{eff}$. It is clearly seen that the effect of superconductivity 
recovering only takes place if the exchange field $h_{eff}$ and the spin accumulation potential $eV$ are very close 
to each other: their difference should be less than $\Delta_0$. Here $\Delta_0$ denotes the value of the 
superconducting order parameter in the film at $h_{eff}=0$. The physical reason for superconductivity
recovering can be easily caught even without solving of a self-consistency equation, already on the level
of consideration of the Cooper's problem of one electron pair. It can be shown that in the presence of 
the exchange field the lowest energy level $\varepsilon_0$ for a pair of electrons with $\bm p_1=-\bm p_2$ 
and opposite spins gets lower upon increasing the spin accumulation potential. Finally, $\varepsilon_0$
becomes exactly equal to its value
for zero exchange field at $eV=h_{eff}$. It is worth to note here that in the absence of the exchange 
field the spin accumulation potential also destroys superconductivity at $|eV| \sim \Delta_0$. 
This effect has been studied in the literature as theoretically so as experimentally 
\cite{takahashi99,vasko97,dong97}.

At $eV=0$ and $T,\Gamma \to 0$ self-consistency equation (\ref{self_con_BCS}) 
has non-zero spatially uniform solution for $h_{eff}<\Delta_0$.
However, it is well-known \cite{larkin64} that the uniform solution become metastable even earlier, 
at $h_{eff}=\Delta_0/\sqrt 2$ (Pauli limiting field) because of the fact that the paramagnetic state is
more energetically favorable for higher exchange fields. In the considered case the paramagnetic state 
cannot be realized because the distribution function is created and supported by the external conditions
in such a way that the populations of majority and minority subbands in the film remain equal. This is 
illustrated in the inset to panel (a) of Fig.~\ref{fig}. The uniform superconducting state is obviously 
more favorable than
the normal one due to the condensation energy. However, there can exist another possibility: in principle,
under spin-dependent quasiparticle distribution an inhomogeneous superconducting state 
(analogous to LOFF-state) can occur and be more favorable than the homogeneous
one at some ranges of parameters. This issue is a prospect for future work. 
    
For the resonance value of $eV_{res}=h_{eff}$ the dependence of the superconducting order parameter on temperature
is very similar to the original BCS one giving practically the same ratio $2\Delta(T=0)/T_c$, as 
illustated in panel (b) of Fig.~\ref{fig}. As it is also represented in panel (b)
of Fig.~\ref{fig}, when $eV$ deviates from $eV_{res}$ the temperature supresses the order parameter more sharply.

Now we turn to the discussion of spin relaxation influence on the effect. We assume spin-flip scattering
from magnetic impurities to be the dominant spin relaxation process inside the superconducting film at
low temperatures. It can be taken into account by adding the corresponding self-energy term 
$\left[-(1/2\pi \tau_{sf})\check {\sigma} \check g \check {\sigma}, \check g \right]$ to the left-hand
side of Eq.~(\ref{usadel}). Here $\check \sigma = [\bm \sigma (1+\tau_3)/2+\bm \sigma^*(1-\tau_3)/2]\rho_0$.
As it is well-known, the influence of the spin-flip scattering is twofold. Firstly, it "works" as a depairing 
factor destroying the coherence peaks and reducing the critical temperature of the superconductor \cite{abrikosov61}.
Secondly, the spin-flip scattering influences directly the distribution function 
reducing the difference $\varphi_\uparrow-\varphi_\downarrow$. The reduction can be roughly estimated as
$\varphi_\uparrow^{sf}-\varphi_\downarrow^{sf}=(\varphi_\uparrow-\varphi_\downarrow)/(1+2\tau_{esc}/\tau_{sf})$. 
Here $\varphi_\uparrow^{sf}-\varphi_\downarrow^{sf}$ is the difference in the presence of spin relaxation processes, 
while $\varphi_\uparrow-\varphi_\downarrow$ is defined by Eq.~(\ref{distrib}) and 
$\tau_{sf}$ is the characteristic spin relaxation time. The results of the exact calculation of the distribution 
function slightly deviate from this rough estimate, especially in 
the region of the coherence peaks. They are represented in panel (c) of Fig.~\ref{fig}. 
Obviously, the "damage" of the distribution function due to spin relaxation also
suppresses the effect of $\Delta$ recovering. This suppression can be roughly viewed as
the effective reduction of the coupling constant $\lambda \to \lambda_{eff}=\lambda(1+\tau_{esc}/\tau_{sf})^{-1}$.
The resulting influence of the spin-flip scattering on the 
order parameter is demonstrated in panel (d) of Fig.~\ref{fig}. 
  
In summary, we have theoretically shown that creation of spin-dependent quasiparticle distribution
in a superconductor can fully compensate the pair-breaking effect of the Zeeman field. Thus, superconductivity
can be recovered for exchange fields well exceeding the Pauli limiting field if the spin accumulation
potential $eV_\downarrow-eV_\uparrow \approx 2h_{eff}$ is generated in the superconductor. It is proposed
that this effect can be experimentaly realized on the basis of voltage biased junction consisting of 
a thin superconducting film sandwiched between two half metals. 

{\it Achnowledgments.} The authors are grateful to V.V. Ryazanov, A.S. Mel'nikov, Ya.V. Fominov and M.A. Silaev
for many useful discussions.



\begin{thebibliography}{99}
%
\bibitem{larkin64}
A.I. Larkin and Yu.N. Ovchinnikov, Sov. Phys. JETP {\bf 20}, 762 (1965) [Zh. Eksp. Teor. Fiz. {\bf 47}, 1136 (1964)].
%
\bibitem{fulde64}
P. Fulde and R.A. Ferrel, Phys.Rev. {\bf 135}, A550 (1964).
%
\bibitem{sarma63}
G. Sarma, J. Phys. Chem. Solids {\bf 24}, 1029 (1963).
%
\bibitem{maki68}
K. Maki, Progr. Theoret. Phys. {\bf 39}, 897 (1968).
%
\bibitem{buzdin05}
A.I. Buzdin, S. Tollis, and J. Cayssol, Phys. Rev. Lett. {\bf 95}, 167003 (2005).
%
\bibitem{bergeret01}
F.S. Bergeret, A.F. Volkov, and K.B. Efetov, Phys. Rev. Lett. {\bf 86}, 3140 (2001).
%
\bibitem{tedrow86}
P.M. Tedrow, J.E. Tkaczyk, and A. Kumar, Phys. Rev. Lett. {\bf 56}, 1746 (1986).
%
\bibitem{meservey94}
R. Meservey and P.M. Tedrow, Phys. Rep. {\bf 238}, 173 (1994).
%
\bibitem{moodera88}
J.S. Moodera, X. Hao, G.A. Gibson, and R. Meservey, Phys. Rev. Lett. {\bf 61}, 637 (1988).
%
\bibitem{hao91}
X. Hao, J.S. Moodera, and R. Meservey, Phys. Rev. Lett. {\bf 67}, 1342 (1991).
%
\bibitem{cottet09}
A. Cottet, D. Huertas-Hernando, W. Belzig, and Yu.V. Nazarov, Phys. Rev. B {\bf 80}, 184511 (2009).
%
\bibitem{bobkova10}
I.V. Bobkova and A.M. Bobkov, Phys. Rev. B {\bf 82}, 024515 (2010).
%
\bibitem{bobkov11}
A.M. Bobkov and I.V. Bobkova, Phys. Rev. B {\bf 84}, 054533 (2010).
%
\bibitem{soulen98}
R.J. Soulen {\it et al.}, Science {\bf 282}, 85 (1998).
%
\bibitem{ji01}
Y. Ji {\it et al.}, Phys. Rev. Lett. {\bf 86},  5585 (2001).
%
\bibitem{park98}
J.-H. Park {\it et al.}, Nature (London) {\bf 392}, 794 (1998).
%
\bibitem{serene83}
J. W. Serene and D. Rainer, Phys. Rep. {\bf 101}, 221 (1983).
%
\bibitem{grein09}
R. Grein {\it et al.}, Phys. Rev. Lett. {\bf 102},  227005 (2009).
%
\bibitem{takahashi99}
S. Takahashi {\it et al.}, Phys. Rev. Lett. {\bf 82}, 3911 (1999).
%
\bibitem{vasko97}
V.A. Vas'ko {\it et al.}, Phys. Rev. Lett. {\bf 78}, 1134 (1997).
%
\bibitem{dong97}
Z.W. Dong {\it et al.}, Appl. Phys. Lett. {\bf 71}, 1718 (1997).
%
\bibitem{abrikosov61}
A.A. Abrikosov and L.P. Gor'kov, Sov. Phys. JETP {\bf 12}, 1243 (1961).
%
\end{thebibliography}

\end{document}